\documentclass[showpacs,twocolumn,aps]{revtex4}
\usepackage{graphicx}
\usepackage{epsfig}
\usepackage{psfrag}

\begin{document}
\title{Theory of orientational ordering in colloidal molecular crystals} 
\author{R. Agra, F. van Wijland and E. Trizac}
\affiliation{Laboratoire de Physique Th\'eorique 
(UMR 8627 du CNRS), B\^atiment 210, 
Universit\'e Paris-Sud, 91405 Orsay, France}
\begin{abstract}
  Freezing of charged colloids on square or
  triangular two-dimensional periodic substrates has been recently
  shown to realize a rich
  variety of orientational orders. We propose a theoretical framework
  to analyze the corresponding structures. A fundamental
  ingredient is that a non spherical charged object in an electrolyte 
  creates a screened 
  electrostatic potential that is anisotropic at any distance.
  Our approach
  is in excellent agreement with the known experimental and numerical 
  results, and explains in simple terms the reentrant orientational melting 
  observed in these so called colloidal molecular crystals. We also investigate
  the case of a rectangular periodic substrate and predict an unusual
  phase transition between orientationnaly ordered states, as the
  aspect ratio of the unit cell is changed.
\end{abstract}
\maketitle


The theoretical scenario put forward by Kosterlitz, Thouless, Halperin, Nelson and 
Young for 2D melting has fostered many experimental investigations \cite{Nelson}.
Most of these studies focused on 2D ordering on homogeneous substrates.
On the other hand, the situation of a corrugated substrate has received
much less attention in spite of its experimental relevance, e.g to describe
an atomic adsorbate on a crystalline surface. Colloidal particles subjected to
light lattices created by laser arrays, appear as an ideal system to 
investigate the latter problem, see Refs \cite{Chow,Chakra,Wei,Bech2001}.
These studies --mostly with 1D periodic substrates-- have revealed 
a novel laser induced freezing and laser induced melting that seems
now well understood \cite{BechFrey}. It was only very recently
that colloidal crystallization on 2D (triangular or square) periodic substrates 
has been investigated both experimentally \cite{Bech2002} and numerically
\cite{RO}. The compatible results reported there call for a theoretical description
that remains hitherto open. A rich variety of novel colloidal states
were reported. When there are more colloids (number $N_c$) than the $N$ substrate 
minima, the $N_c/N$ colloids trapped in each well may be considered
as a single bound entity with only rotational degrees of freedom. These
``colloidal molecules'' (dimers for $N_c=2N$, trimers for $N_c=3N$\ldots) 
were found experimentally and numerically to
display long range orientational order. The corresponding state has
been coined ``colloidal molecular crystal'' in Ref. \cite{RO}. Quite 
an unusual 
two stage melting process may be observed for such systems. Decreasing the
temperature $T$ from an initial standard 2D liquid, individual diffusion
is first inhibited. The particles localize in the substrate wells
(traps) thereby forming the colloidal molecules. The orientational order
in this partially ordered solid is short range. Upon further decreasing $T$,
the colloidal molecules no longer rotate freely and acquire orientational order. 
Surprisingly, increasing the substrate strength $V_0$ at fixed $T$ leads
to reentrant melting: the colloidal molecular crystal is destabilized 
into the partially ordered solid, and orientational order is lost.

The aim of the present Letter is to provide a simple theoretical framework
for the phase behavior of such a system, that is very different from that 
on homogeneous or 1D periodic substrate potentials. We adopt here the
simplest viewpoint and consider a trapped molecule ($n$-mer) as a single rigid
object with $n$-fold symmetry and one orientational degree of freedom within 
a common given plane. 
Additional degrees of freedom such as radial or angular
fluctuations will be discarded. We shall consequently not investigate the
liquid phase (at high $T$ and low $V_0$ where the $n$-mers are dissolved)
and concentrate on the orientational freezing transition.
In connection with numerical and experimental outputs,
three questions naturally arise:
a) is it possible to predict the non trivial ground state observed 
in the experiments and simulations? b) what are the 
characteristics of the above phase transition, in particular, its order?
c) can we explain the reentrant melting considering rotational 
degrees of freedom only? In addition, we provide predictions 
for further experimentally relevant situations.

\begin{figure}[hb]
$$\epsfig{figure=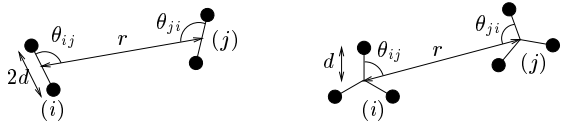,width=8.cm,angle=0}$$
\caption{\label{fig:Fig1} Definition of the notations adopted to compute 
dimer (left) and trimer (right) interactions. The black dots show the
colloids, and the line between them is a guide to the eye.}
\end{figure}

Before addressing these questions, general considerations concerning 
screened electrostatic interactions between anisotropic charged 
molecules (the $n$-mers) are in order. Because their surface groups dissociate
in solution, the colloids under scrutiny here are highly charged. They strongly 
repel each other and experience a screened Coulomb (Yukawa) potential 
\cite{DLVO} $\phi(r) \propto \exp(-\kappa r)/r$, where the $r$ dependence
only is relevant for the discussion, not the prefactor depending both on
particle size and charge \cite{RO}. In this expression, $r$ is the center to center
distance and $\kappa^{-1}$ the Debye screening length (smaller than the 
periodicity of the underlying 2D substrate). Once an $n$-mer is formed
in a trap, how does it interact with a neighboring $n$-mer? Consider
dimers for simplicity. One could naively think that in an electrolyte,
the electrostatic potential created by such an object becomes isotropic at large
distances, as is the case in vacuum or in a plain dielectric medium. 
This is however incorrect and summing the screened contributions
from each particle forming dimers $i$ and $j$, 
one obtains the leading order large distance expression
(see Fig. \ref{fig:Fig1} for the notations used)
\begin{equation}
\phi_{ij}(r,\theta_{ij},\theta_{ji}) \propto \cosh[\kappa d \cos(\theta_{ij})] 
\cosh[\kappa d \cos(\theta_{ji})] \, \frac{e^{-\kappa r}}{r}
\label{eq:potdim}
\end{equation}
The key point here is that the radial and angular dependence of the interaction
potential are factorized. The anisotropy of a dimer is therefore felt at all
distances even for $\kappa r \gg 1$, provided the Debye length is small 
enough to resolve the dimer structure (when $d \ll 1/\kappa$, the 
isotropic screened Coulomb potential is recovered). It is noteworthy to mention
in passing that considering dipoles (two point charges
$+q$/$-q$ at a distance $2d$) instead of a dimer $+q$/$+q$,
the same potential would hold provided the cosh function in Eq. 
(\ref{eq:potdim}) is replaced by a sinh. The angular dependence of such 
an expression differs much from that of the unscreened 
dipolar interaction. In addition, the distance dependences of both 
dimer and dipole potentials are identical ($\exp(-\kappa r)/r$ in the far field
region). The same holds for higher order multipoles, that all contribute
to the same order in distance. As a consequence, when considering 
the interactions between arbitrary charged objects in a electrolyte,
a usual truncated multipolar-like expansion \cite{Jackson} is doomed to fail;
in the case of dimers, resuming all multipole orders would lead to
Eq. (\ref{eq:potdim}), to leading order in $r$.
These considerations, that seem to have been overlooked in soft matter
literature, appear to account in simple terms for orientational freezing in
colloidal molecular crystals. 

\begin{figure}[ht]
\epsfig{figure=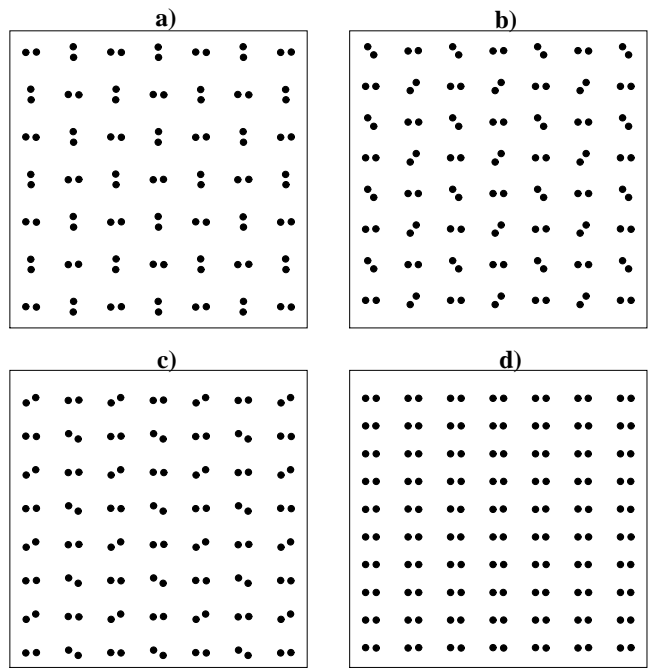,width=9cm,angle=-90}
\caption{Configuration of colloids (black dots) at $T=0$
on a rectangular lattice of traps, when $N_c=2N$ and $\kappa d=1.5$. 
a) situation on 
the square lattice (aspect ratio $\alpha=1$); b) rectangular lattice
with $\alpha=0.90$; c) $\alpha=0.85$; d) $\alpha=0.7$ i.e. smaller than
the threshold $\alpha^*$ reported in Fig \ref{fig:Fig3}.
Plots b) and c) correspond to the transition region $\alpha^*<\alpha<\alpha^{**}$
between F and AF phases.}
\label{fig:Fig2}
\end{figure}

In Ref. \cite{RO}, Reichhardt and Olson showed numerically that dimers on a 
square 2D periodic substrate (hereafter referred to as the underlying lattice
of traps) adopt the ground state represented in Fig. \ref{fig:Fig2}-a).
Such an ordering should follow from minimizing the total 
electrostatic energy $\cal E$ where each pair of dimers interact 
through the potential (\ref{eq:potdim}). Given the exponential
screening with distance and that $\kappa^{-1}$ is smaller than lattice
spacing, we further restrict to nearest neighbors of dimers and write the 
relevant angular dependent part of the energy as
\begin{equation}
{\cal E} = \sum_{\langle i,j\rangle} \cosh[\kappa d \cos(\theta_{ij})] 
\cosh[\kappa d \cos(\theta_{ji})],
\label{eq:calE}
\end{equation}
where the angular brackets denote nearest neighbors.
A given pair minimizes its repulsion by setting
$\theta_{ij}=\theta_{ji}=\pi/2$ (parallel dimers, perpendicular
to the center to center vector, see Fig. \ref{fig:Fig1}).
Such a pair configuration is however not space-filling,
so that no trivial ground state may be identified at this stage. 
By analogy with spin systems, 
we may invoke frustration. To proceed further, it is instructive to
approximate the cosh function in (\ref{eq:calE}) by a parabola. Introducing 
next a new set of variables where 
$\theta_{i} \in [-\pi/2,\pi/2]$ is defined as the 
angle between dimer axis and one of the principal lattice directions
\cite{rque}
we get 
\begin{equation}
{\cal E} \,=\, \hbox{cst}\, +\, (\kappa d)^4 \,\sum_{\langle i,j\rangle}
\left(\cos\theta_i\right)^2 \,\left(\cos\theta_j\right)^2.
\end{equation}
Remarkably, this expression takes the form of an antiferromagnetic
Ising-like Hamiltonian, once the variables $\sigma_i = \cos2\theta_i$
have been introduced ($-1\leq \sigma_i\leq 1$):
\begin{equation}
{\cal E} \,=\, \hbox{cst} \,+\, (\kappa d)^4 \,\sum_{\langle i,j\rangle}
\sigma_i \sigma_j .
\label{eq:simple}
\end{equation}
With the ``spin'' variables, frustration vanishes
and the ground state simply consists of $\sigma_i=\pm 1$,
with opposite signs on neighboring sites. This corresponds
to $\theta_i=0$ or $\pi/2$, which is in perfect agreement with the results
of the numerical simulations of Ref. \cite{RO} (see Fig. \ref{fig:Fig2}).
We have checked that keeping the original expression 
(\ref{eq:calE}) without the parabolic approximation for the cosh leads to the
same results.

Having answered question a) above in the case of dimers on the square lattice, 
we now address questions b) and c),
in particular the reentrant melting phenomenon.
At high temperature, the system has to be in a disordered state,
a paramagnetic phase in spin language,
and on purely dimensional grounds, Eq (\ref{eq:simple}) indicates that 
the critical temperature of the antiferromagnetic/paramagnetic transition
scales like $T_c\propto (\kappa d)^4$. At fixed temperature $T$,
increasing the substrate pinning strength $V_0$, the $n$-mers constituting a
colloidal molecule become of smaller spatial extent: $d$ decreases,
so that $T_c$ eventually becomes smaller than $T$. Orientational
order  is correspondingly lost.
More specifically, we find
$T_c \propto V_0^{-4/3}$.
As emphasized above, we discard the possible loss of positional order,
that occurs at large $T$. The corresponding 
fusion temperature $T_f$ has to increase with $V_0$. For large enough $V_0$,
the transition between the ordered and partially ordered solids
is therefore always present, since $T_c < T_f$. At low $V_0$,
this transition is preempted by the positional melting $(T_f<T_c)$.

The above arguments may be rationalized by computing the partition function 
associated to the Hamiltonian (\ref{eq:calE}) \cite{rque3}. A mean-field analysis,
simplified by the introduction of
two sub-lattices to describe the order displayed in Fig. \ref{fig:Fig1},
leads to the conclusion that orientational freezing is a second order phase 
transition \cite{toappear}. 

The same procedure may be extended
to other substrate geometries, or more complex colloidal molecules.
For $n$-mers, the counterpart of Eq. (\ref{eq:potdim}) reads
\begin{equation}
\phi_{ij}(r,\theta_{ij},\theta_{ji}) \,=\,
f(\theta_{ij},\kappa d) f(\theta_{ji},\kappa d)\,\frac{e^{-\kappa r}}{r}
\end{equation}
where
\begin{equation}
f(\theta,\kappa d) \, = \,
\sum_{k=0}^{n-1} \exp\left[\kappa d\cos\left(\theta+\frac{2k \pi}{n}\right)\right],
\end{equation}
which reduces to Eq. (\ref{eq:potdim}) in the case of dimers ($n=2$).
The variable $\theta_{ij}$ is defined as the angle between one arm of the symmetric
$n$-fold colloidal molecule $i$ and the center to center direction $(ij)$,
see Fig. \ref{fig:Fig1} for the cases $n=2$ and $n=3$. 
For the situation of a triangular substrate with trimers 
as colloidal molecules (the experimental situation 
investigated in \cite{Bech2002}), we may
follow similar steps 
as above, i.e. minimize $\sum f(\theta_{ij})f(\theta_{ji})$ where the
summation is restricted to nearest neighbors. This leads to the
ground state where all trimers point in the same direction,
with an angle $\theta=\pi/3$ between one arm of the trimers and one
of the principal lattice direction (or equivalently, 
$\theta_{ij}= \pi/3\pm \pi/3$ 
depending on the pair ($i$-$j$)). This is exactly the order observed
experimentally \cite{Bech2002}, and in numerical simulations \cite{RO}.
For dimers on a triangular substrate (not studied experimentally), 
we also obtain the same
orientational order as in Ref. \cite{RO} (herringbone structure with
$\theta=\pm \pi/4$, or equivalently 
$\theta_{ij} = \pi/4 \pm \pi/3$).

Our framework allows to investigate the situation of non-integer filling.
We will focus on the half integer case $N_c/N=3/2$, as considered numerically
in \cite{half-integer}. To minimize electrostatic repulsion on the square
lattice, the colloids create a checkerboard order with two
sub-lattices such that the nearest neighbors of monomer are
exclusively dimers, and vice-versa \cite{half-integer}. In this 
configuration, nearest neighbors interaction do not depend on the
angular coordinates, so that the electrostatic energy to
consider should include second nearest neighbor terms. We find that 
the corresponding ground state is degenerate, with dimer
orientations $\theta_i=0$ or $\theta_i=\pi/2$. Since there are
$N_c$ dimers, this ground state is $2^{N_c}$ degenerate,
and no long range order can be observed.
This finding supports the numerical results reported in \cite{half-integer}.
Changing the geometry of the underlying substrate from  square to
triangular does not allow to tile the lattice such that minima
with only one colloid will be in between every other dimer
\cite{half-integer}. The
partitioning of the $3N_c$ colloids in the $2N$ traps is therefore
a frustrated problem with no ordering, be it positional or orientational.

\begin{figure}[ht]
\epsfig{figure=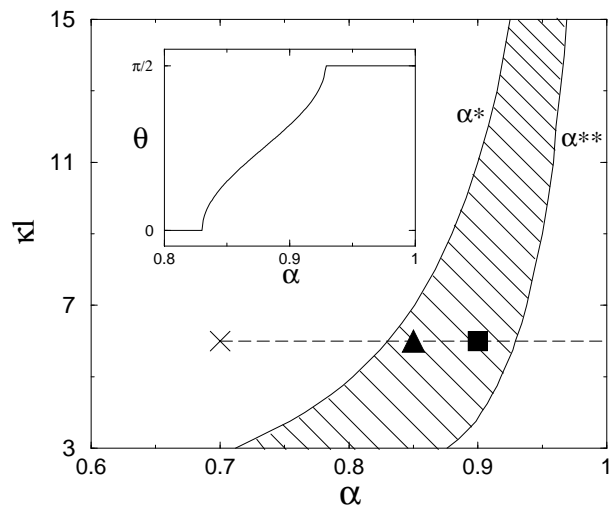,width=8cm,angle=0}
\caption{ Phase diagram for dimers 
on a rectangular substrate as a function of the aspect ratio
$\alpha$. The F phase is obtained for $\alpha<\alpha^{*}$,
while $\alpha>\alpha^{**}$ corresponds to the AF region.
Here $\kappa d$ is fixed to 1.5,
which would correspond experimentally to a fixed depth $V_0$.
For a given value $\kappa l=6$ (horizontal dashed line in the
main graph such that $\alpha^*\simeq 0.82$ and $\alpha^{**}\simeq 0.93$), 
the inset shows how the system evolves from 
the F to AF phase as $\alpha$ increases. 
The cross, triangle and square correspond respectively to the situations of
Fig. \ref{fig:Fig2}-d) with $\alpha=0.7$, 
Fig. \ref{fig:Fig2}-c) with $\alpha=0.85$, 
and Fig. \ref{fig:Fig2}-b) with $\alpha=0.90$.
 }
\label{fig:Fig3}
\end{figure}

We now consider a final geometry of experimental
relevance, namely dimers on a rectangular
lattice (with unit cell of lengths $\alpha l$ and $l$; by definition,
$\alpha<1$, and the direction along the long axis is called 
``horizontal'' for convenience). When the aspect
ratio $\alpha$ is 1 the checkerboard order is that of Fig.
\ref{fig:Fig2}-a) (antiferromagnetic-like phase, denoted AF). On the other
hand, when $\alpha\ll 1$, there are only two nearest neighbors 
per particle, and from expression (\ref{eq:potdim})
the preferred configuration is $\theta_{ij}=\theta_{ji}=\pi/2$:
all dimers are parallel to the horizontal direction. This
ferromagnetic-like phase is denoted F. We have studied the transition 
between the F and AF phases, minimizing the relevant energy
\begin{equation}
{\cal E}_{\text{rect}} \,=\, \sum_{\langle i,j\rangle} \beta_{ij}
\cosh[\kappa d \cos(\theta_{ij})] \cosh[\kappa d \cos(\theta_{ji})],
\end{equation}
where $\beta_{ij} = \alpha^{-1} \exp[\kappa l (1-\alpha)]$  
(resp. $\beta_{ij} =1$) for a vertical (resp. horizontal) bond $i$-$j$.
As expected, when $\alpha$ is close to 1 (in fact $\alpha>\alpha^{**}$), 
the order is the AF phase
of Fig. \ref{fig:Fig2}-a) and when $\alpha$ is small enough
($\alpha<\alpha^*$), the 
F phase of Fig. \ref{fig:Fig2}-d) is observed.
As shown in Fig. \ref{fig:Fig3}, the thresholds $\alpha^*$ and $\alpha^{**}$
depend on screening conditions and dimer length 
through $\kappa l$ and $\kappa d$. Investigating the behavior 
in the transition region 
$\alpha^*<\alpha<\alpha^{**}$ requires the inclusion of next nearest
neighbors \cite{toappear}, and leads to the conclusion that 
the system divides into two sub-lattices A and B 
(see Fig \ref{fig:Fig2}-b) and c)). On A, the dimers remain parallel to 
the horizontal axis while on B, they have alternating orientations
$\pm\theta$ depending on the row (see Fig. \ref{fig:Fig2}). 
The inset of
Fig \ref{fig:Fig3} shows how the orientations evolve from
the F state with $\theta=0$ to the AF order with $\theta=\pi/2$.
The F/AF transition driven by $\alpha$ is therefore second order.
We have also studied more complex situations with higher 
fillings (integer and half integer), that realize a  
host of non trivial colloidal crystalline states \cite{toappear}.

Our approach relies on the validity of the pair potential
(\ref{eq:potdim}) which is an expansion obtained at large distances
from the original Yukawa expression. Such an expansion is
justified for $d\ll  l$ and $(\kappa d)^2 \ll \kappa l$. In addition,
we restricted to nearest or second nearest neighbors 
interactions, which requires $\kappa l \gg 1$. Knowledge of the
precise parameter bounds where our predictions apply --in 
particular on the triangular lattice-- is an 
important question of experimental relevance, that can only be answered
numerically. Equally relevant is the influence of temperature,
that may affect the scenarios reported here for the ground
state behaviour (such as those of Figs \ref{fig:Fig2} and
\ref{fig:Fig3}). Work along these lines is in progress,
with Monte Carlo simulations.

In conclusion, our theoretical approach explains in simple terms
the rich orientational phenomenology of colloidal molecular crystals,
both the emergence of long range order and reentrance.
Our work establishes a natural connection between 
the situation of dimers on a square lattice and
the canonical Ising model of statistical mechanics.
The possibility of such a mapping however seems lost
for more complex colloidal molecules or with other substrate
geometries. The mechanisms at work here discard 
the radial and angular fluctuations of a trapped colloidal 
molecule, that appear to be of secondary importance. 
On the other hand, such fluctuations are prevalent on one-dimensional
periodic substrates \cite{BechFrey} and could {\it a priori}\/ 
have played a role in the present two-dimensional situation \cite{Bech2002}.
Finally, we hope that our results will stimulate further experimental
investigations in this field.


We would like to thank C. Bechinger, S. Bleil, L. Bocquet, M. Brunner,
B. Jancovici and Z. R\'acz for useful discussions.


\end{document}